\documentclass{article}
\usepackage{cite}
\usepackage{amsmath,amssymb,amsfonts}
\usepackage{algorithmic}
\usepackage{spconf, graphicx}
\usepackage{textcomp}
\usepackage{xcolor}
\usepackage{float}
\usepackage{booktabs}

\usepackage[draft]{changes}
\setaddedmarkup{\color{blue}{#1}}
\setdeletedmarkup{\color{orange}{\sout{#1}}}

\begin{document}

\title{Early prediction of the transferability of bovine embryos \\ from videomicroscopy
}

\twoauthors
  {Y. Hachani$^1$, P. Bouthemy$^1$, E. Fromont$^{2}$ 
 }
	{$^1$Inria, 
     $^{2}$Univ. Rennes, IUF, Inria, IRISA\\
	France}
 {S. Ruffini$^{4}$, L. Laffont$^{4}$, A. De Paula Reis$^{3,4}$  
  }
{$^{3}$Ecole Nationale Vétérinaire d’Alfort \\
 $^{4}$University Paris-Saclay, UVSQ,
   INRAE, BREED\\
    France}

\maketitle

\begin{abstract}
Videomicroscopy is a promising tool combined with machine learning for studying the early development of \textit{in vitro} fertilized bovine embryos and assessing its transferability as soon as possible. We aim to predict the embryo transferability within four days at most, taking 2D time-lapse microscopy videos as input. We formulate this problem as a supervised binary classification problem for the classes \emph{transferable} and \emph{not transferable}. The challenges are three-fold: 1) poorly discriminating appearance and motion, 2) class ambiguity, 3) small amount of annotated data. We propose a 3D convolutional neural network involving three pathways, which makes it multi-scale in time and able to handle appearance and motion in different ways. For training, we retain the focal loss. Our model, named SFR, compares favorably to other methods. Experiments demonstrate its effectiveness and accuracy for our challenging biological task. 
\end{abstract}


\section{Introduction}

\begin{figure}[bh!]
\centering
\resizebox{\columnwidth}{!}{
        \includegraphics{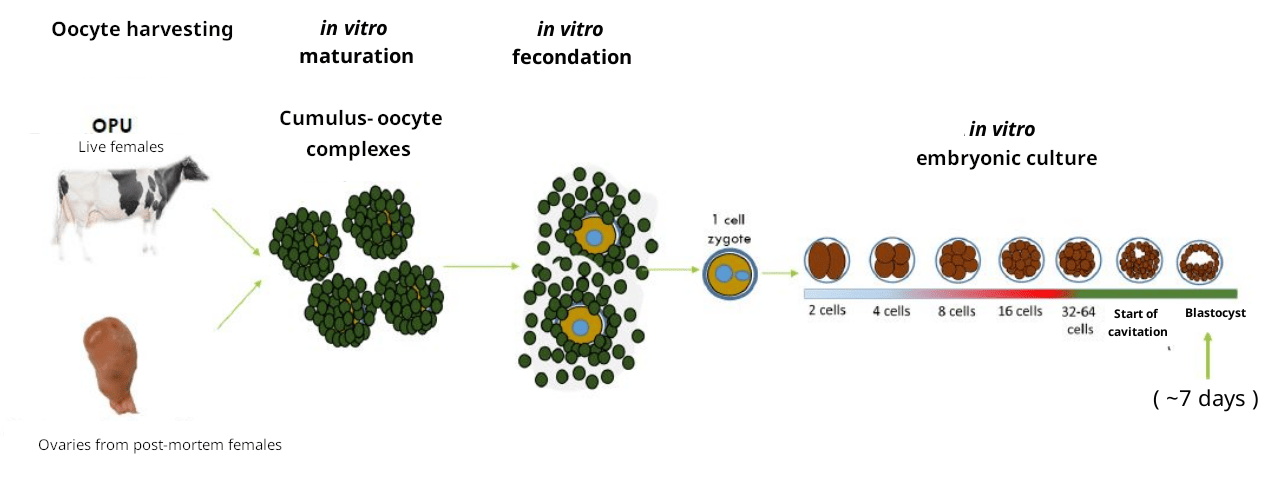}}
    \caption{The \textit{in vitro}  fertilization (IVF) process for bovine embryos.}
    \label{fig-IVF}
\end{figure}

Most techniques used to study the mechanisms of embryonic development are incompatible with embryo survival. Video microscopy applied to bovine embryos produced by \textit{in vitro} fertilization (IVF) (Fig.\ref{fig-IVF}), is a promising tool compatible, with survival, in association with the analysis power of machine learning techniques. It allows us to study the early development and to assess the transferability of \textit{in vitro} fertilized embryos, i.e., the capacity to reach the blastocyst stage, suitable for transfer to a cow uterus. From the application point of view, having this ability to correctly predict on a large scale whether embryos can be transferred or not, is crucial for cattle breeding. With current methods, only 30\% of transferred blastocysts actually result in pregnancy. This achievement should help reduce pregnancy failures and thus unnecessary inseminations.

However, two major problems arise: \textit{i)} detailed analysis of each video is time-consuming for biologists and potentially limits a day-to-day practice, \textit{ii)} to enable advanced biological studies on early mechanisms influencing embryo development, it is preferable to know the embryo transferability as soon as possible. Therefore, automating the transferability prediction by operating directly on videos of embryonic development is of key interest, while achieving the task the earliest possible.

\begin{figure}[tbh!]
    \centering
    \includegraphics[width=0.7\columnwidth]{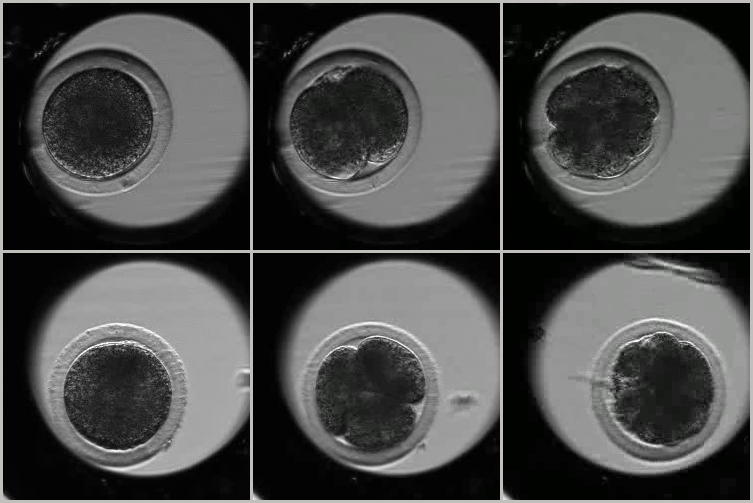}
    \caption{Two sample videos of IVF bovine embryos, with three images taken at distant time instants (top row: an example from class transferable; bottom row: an example from class non transferable). The bovine embryo (dark grey), surrounded by the zona pellucida, is located in a micro-well (light grey) within the Petri dish (black). The embryo occupies a small part of the image. }
    \label{embryo-dev}
\end{figure}

Our overall objective is then to achieve a correct prediction of the embryo transferability within four days at most, taking 2D time-lapse videos as input to be analyzed by 3D convolutional neural networks. This achievement will offer biologists new prospects such as better embryo sorting for further studies on embryonic genome activation (EGA) (at day 4) or morula formation (at day 5), decision-taking for earlier transfer into the uterus of a recipient female, which limits the duration of culture under sub-optimal conditions. 

We formulate this problem as a two-class supervised classification one: the embryo is transferable (T class) or not transferable (NT class). This problem is challenging for three main reasons: 1) tricky embryo appearance and motion, 2) class ambiguity, and 3) small amount of annotated data. First, as illustrated in Fig.\ref{embryo-dev}, the microscopy videos display little contrast, involve a lot of noise and complex motion with transparency effects. The diverse embryos studied often exhibit poorly discriminating appearance between classes (see Fig.\ref{embryo-dev} again), while the videos show complex morphological and temporal processes. Second, the intra-class variability is high, in the sense that the trajectories of the embryo development may substantially vary within a given class. Conversely, the inter-class distance is low. Indeed, observed development of two embryos from the two different classes may be fairly similar for the four first days. Third, because labeling is costly, there is only a limited database of videos labeled transferable or not transferable. 

The rest of the paper is organized as follows. Section \ref{rw} describes related work. In Section \ref{method}, we present our 3D network with three pathways, and the associated training losses. Section \ref{results} reports quantitative and comparative evaluations of the method. Section \ref{conclusion} contains concluding remarks.

\section{Related work}
\label{rw}

Biologists have been working on embryonic development and phenotype formation for several decades. Thanks to IVF, significant advances have been achieved leading to major progress in medicine and breeding. Video-microscopy for bovine embryos has enabled biologists to observe different development trajectories leading to distinct phenotypes.

In human artificial reproductive medicine, the objective is to reduce the need for multiple pregnancies and losses during pregnancy. In \cite{human_embryo0}, a first human embryo selection model was introduced based on manual annotation and a decision tree. Then, the authors of \cite{human_embryo1} developed an embryo selection model using a single static image captured by light microscopy. If the results of these studies were promising, the learning process was carried out \textit{retrospectively}. Indeed, the pregnancy outcomes were predicted from embryos that were previously manually selected by an expert for transfer, leaving a large number of embryos outside the studies. These results may therefore include a certain level of over-fitting, as well as a number of biases related to embryo manipulation and transfer. Recently, a deep learning approach has been adopted in \cite{human_embryo2} to select human embryos from time-lapse image sequences acquired over five days, knowing that the video acquisition started around 24 hours after insemination. They used the Inflated 3D ConvNet (I3D) of \cite{kinetics} followed by a recurrent LTSM network. However, training is still carried out retrospectively. It leverages a very large dataset (about 100,000 videos).

Deep learning (DL) has also been used on human embryo videos to tackle different problems.  For instance, in \cite{jang2023}, the authors proposed a vector quantized variational autoencoder (VQ-VAE) to segment blastomere instance. Latest work mostly focuses on characterizing the different stages of embryo development, namely cell cleavage with intermediate stages defined by cell count (from 1-cell to 4-cell, sometimes up to 8-cell count), morula and blastocyst. In \cite{stage_detection_synergic}, the authors elaborated a development stage detector based on a 2D-CNN followed by a LSTM network classifier and added a synergic loss to learn embryo-independent features. In \cite{embryosformer}, the development stage classification was improved with EmbryosFormer, a three-headed model designed as an encoder-decoder deformable transformer inspired from Deformable DETR\cite{def_detr}. The authors of \cite{stage_detection_yolo} adopted a different approach using the object detection technique YOLO v5 \cite{yolov5} and performing cell counting.
  
Bovine embryos are more difficult to study than their human counterparts, since their cells are darker, which makes, for example, cell counting quite difficult. Besides, biologists have reported \cite{bovine_embryo0} that the development trajectories may reflect different adaptation mechanisms and aptitudes for future gestation. In \cite{bovine_embryo0}, the authors proceeded \textit{prospectively}: they first characterized the trajectories and verified their biological interest, limiting the biases of the retrospective studies mentioned above. Observations based on the embryonic morphokinetic features have led to distinguish several families of trajectories distributed in transferable embryos (here, named T class) and non-transferable embryos (here, named NT-class). A prediction model was defined, based on random forest, leveraging many detailed annotations for each video.

Embryonic development could be seen as a form of action (in the computer vision terminology), and then, our classification problem could be seen as an action classification one. We therefore briefly review work on action recognition in videos since the advent of deep learning. The pioneering work \cite{two-streams} introduced a two-stream convolutional network taking both images and optical flow fields as input to leverage appearance and motion for action recognition. However, we experienced that optical flow was poorly estimated on bovine embryo videos. By considering the spatio-temporal video as a 3D volume, 3D convolutional networks have been extensively adopted since then for action recognition, as shown for instance in \cite{kinetics} with the inflated 3D convnet or in \cite{resnet} with a 3D ResNet. In \cite{slowfast}, the authors proposed a 3D network comprising two pathways, a Slow one devoted to appearance information with input video at a low frame rate, and a Fast one with input video at high frame rate to better capture motion. This last model will be inspirational in our work. 

 To the best of our knowledge, the model we propose is the first DL-based model devoted to bovine embryos that are more difficult to handle. In addition, we focus on the transferability of IVF embryo in a prospective way, and the earliness of the prediction is an essential aspect of our work. Consequently, we consider a shorter period of embryonic development. Finally, we only leverage one annotation per video, that is, its class, transferable or not transferable, and a limited amount of annotated videos.

\section{Model description}
\label{method}

As stated above, we address a two-class classification problem to predict the transferability of the IVF bovine embryos. The two classes are transferable embryos (T class - embryos with a potential of establishing a pregnancy, they can be transferred into a female recipient) \textit{vs} non-transferable embryos (NT class - embryos with no or very low potential of establishing a pregnancy, they should not be transferred).

In practice, the expert biologist annotates the videos on a longer temporal basis than 4 days, to up to six or even eight days of the embryo development. This is nevertheless a light annotation, one label (the class) per video. In some cases, the expert biologist may need to understand the whole evolution of the embryo over the full video
to decide on the class, as development trajectories may be similar up to a certain stage. On our side, we take these annotations for the same videos but restricted to four days of development. This explains why the inter-class distance may be small when considering videos of 4-day development only. Consequently, automatically predicting transferability at four days, i.e., carrying out our binary classification, is a complex video analysis task. In addition, only a small set of annotated videos is available, the images are noisy and poorly contrasted, subject to transparency effects, and motions in the video are not easy to identify.

\subsection{Network architecture}

We have designed a 3D network for our two-class classification problem in order to achieve early prediction of IVF bovine embryo transferability.
We believe
that a 3D convolutional network is more adapted to properly capture the spatio-temporal features characterizing the embryonic development, than for instance a recurrent neural network. Indeed, the morphokinetic features are quite intricate and an embryo development is not as smooth along time as a human action in a video. It is mainly specified by a few discrete events corresponding to the cell divisions, with rather random local motions in between.

\begin{figure*}[h!]
\centering
    \includegraphics[width=130mm]{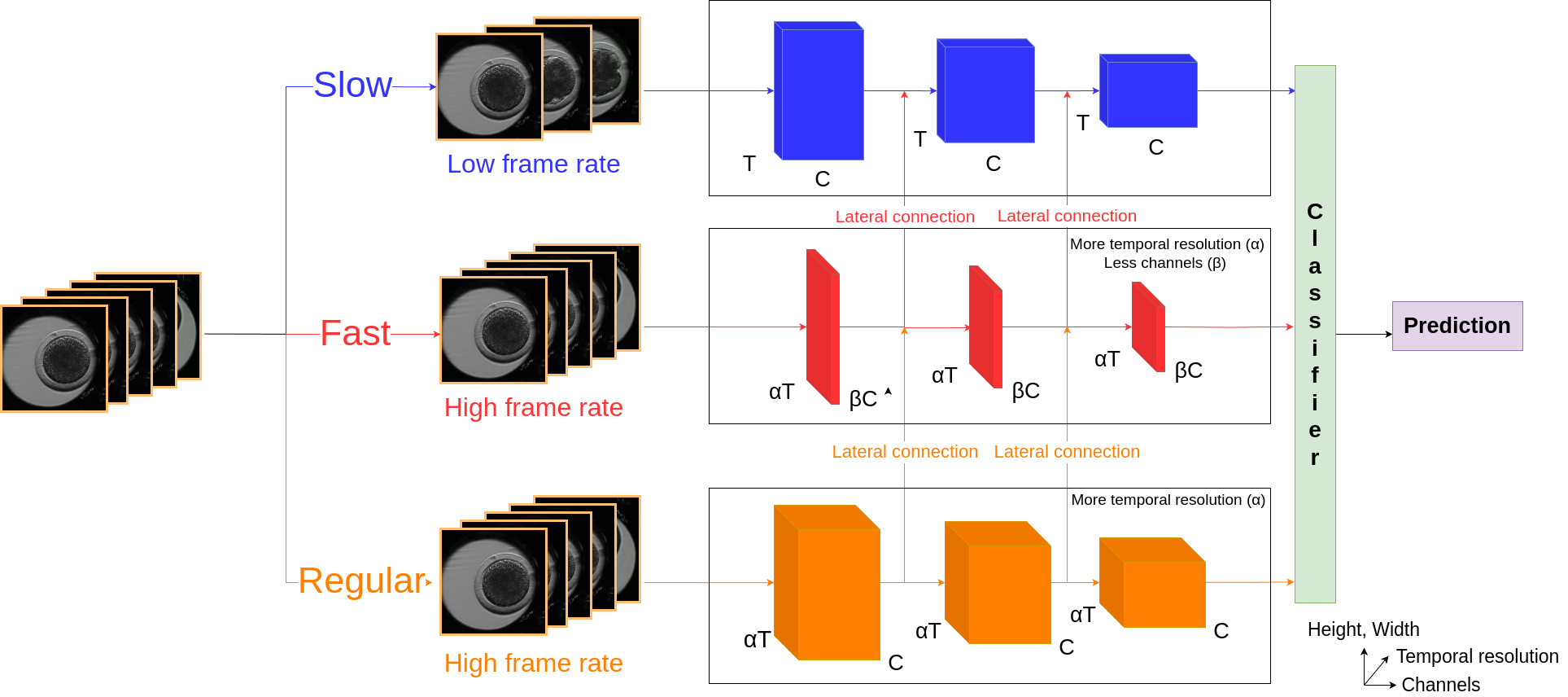}
    \caption{Our 3D model combines three pathways, \textit{Slow} and \textit{Fast} and \textit{Regular}, all implemented with 3D ResNet18 of different configurations. Input video is given at a lower temporal rate for the \textit{Slow} pathway. The three outputs are combined before providing the prediction. Our SFR model includes directed lateral connections between the pathways as drawn in the figure. 
    }
    \label{fig:3Dmodel}
\end{figure*}

Our 3D network presented in Fig.\ref{fig:3Dmodel} involves three pathways combined, with directed lateral connections.
As in the SlowFast network \cite{slowfast}, we have the \textit{Slow} pathway taking the input video at a low frame rate and mainly dedicated to capture spatial features in images, the \textit{Fast} pathway with input video at a high frame rate mainly devoted to the temporal features. The \textit{Fast} pathway has a fraction $\beta$ of channels and a temporal resolution $\alpha$ times higher than the \textit{Slow} pathway.
We call the third pathway \textit{Regular}. It takes the input video at the same rate as \textit{Fast} pathway, but it involves more channels in each layer. As motivated in the ablation study (Section \ref{ablation-study}), we use ResNet18 for the three pathways to build a light 3D network, which speeds up training and mitigates over-fitting.  We replaced all ResNet batch normalization layers with group normalization layers \cite{groupnorm}, since group normalization is at least as good as batch normalization when trained with small or medium batch sizes, and it allows us to use Pytorch Lightning gradient accumulation technique efficiently.

We found the use of the three pathways beneficial, because appearance and motion are rather intertwined due to the transparency of cell membranes and the fact that we observe 2D projections, partly overlaid, of 3D cells. The three pathways bring complementary ways of handling appearance and motion to provide the right prediction. The outputs delivered by the last layer of each pathway are concatenated to feed the classifier. Directed lateral connections are included between pathways,
 We focused on two combinations of the lateral connections.
The first one is illustrated in Fig.\ref{fig:3Dmodel} and comprises a connection from the \textit{Regular} to the \textit{Fast} pathways and from the \textit{Fast} to the \textit{Slow} pathways. The second one involves a fusion from the \textit{Regular} to the \textit{Slow} pathways and from the \textit{Fast} to the \textit{Slow} pathways; the \textit{Regular} and the \textit{Fast} pathways are not connected. We selected the first combination as explained in the ablation study (Section \ref{ablation-study}). We call our method SFR.




\subsection{Loss function}
We could adopt different loss functions. Since our data are unbalanced between the two classes \textit{T} and \textit{NT}, we have considered the focal loss \cite{focal}, initially introduced for the object detection task. The focal loss can contribute to correct this imbalance, while focusing on the most difficult examples. The focal loss writes:
\begin{equation}
\mathcal{L}_f(\textbf{v},y) = -\sum_{c=1}^{2}\alpha_c(1-\hat{p}(y_c|\textbf{v}))^\gamma p(y_c|\textbf{v})\log\hat{p}(y_c|\textbf{v}),
\end{equation}
where $\bf{v}$ denotes the video input, $y_c$ one of the two classes, $\hat{p}(y_c|\textbf{v})$ the predicted probability of having class $c$ given video $\bf{v}$, and $p(y_c|\textbf{v})$ the true one, equal to $1$ for the right class $c$ regarding $\bf{v}$ since we are dealing with supervised classification. In addition, $\alpha_c$ is the weight for class $c$, $\gamma$ the focusing parameter. The larger $\gamma$, the less importance is put to well-classified samples.

As reported in the ablation study, we also investigated the cross-entropy loss \cite{yao2019}, defined for a binary classification by:
\begin{equation}
\mathcal{L}_{ce}(\textbf{v},y) = -p(y_c|\textbf{v})\log \hat{p}(y_c|\textbf{v}) - (1-p(y_c|\textbf{v}))\log(1-\hat{p}(y_c|\textbf{v})).
\end{equation} 



\subsection{Data augmentation}
Various strategies can be adopted to deal with the lack of data. Data augmentation is a classical one \cite{shorten2019}. Here, we can consider data augmentation applied to photometric, spatial, or temporal features of the video \cite{sslv}. In practice, we only applied basic image manipulations to every frame of the videos: Gaussian noise addition, Gaussian blur, image flipping, image transpose, image cropping. All images in a given sequence are modified in the same way. We have thus multiplied the total number of videos in the training set by 20.


\section{Experimental results}
\label{results}
\subsection{Video capture and video dataset}
The videos are acquired as follows. The oocytes recovered from slaughterhouse ovaries and matured \textit{in vitro} are brought into contact with frozen-thawed semen in a culture dish defining the starting point of the biological development of the embryos
\cite{bovine_embryo0}. The embryos are put in microwells of Petri dishes around twenty-two hours after the \textit{in vitro} fertilization. Each Petri dish contains sixteen microwells. It is placed into the PrimoVision system that comprises a simple transmission light microscope, and is filmed by the incubator camera.

The PrimoVision system
takes a picture of the Petri dishes every fifteen minutes over eight days, and delivers 2D time-lapse video sequences that are subsequently divided into sixteen videos, one per embryo. Each video is annotated by a biologist with the \textit{T}-label or the \textit{NT}-label. For the purpose of our work, we have retained only the video footage of the first four days of embryo development. Since video acquisition begins only 22 hours after IVF, each processed video covers a period of 3 days and comprises around 300 images.
The video dataset includes 947 videos, distributed into 763 for training and validation and 184 for test. Each set includes around 65\% of non-transferable embryo videos.


\subsection{Implementation details}

Each model was trained using the AdamW optimizer \cite{adamw}, with a learning rate of $10^{-4}$ and the other parameters kept at their default values.  We applied a cyclic learning rate scheduler as recommended in \cite{cycliclr}. We trained the models using mini-batches of 32 samples artificially created thanks to the accumulate gradients technique, implemented in PyTorch Lightning, that accumulates gradients of small batches before performing a backward pass. We apply the stochastic weight averaging (SWA)\cite{swa} technique, which improves the generalization of our models by averaging the network weights at different, well-chosen epochs. We use early stopping to end training when the loss computed on the validation set increased ten epochs in a row. Then, we select the model at the epoch with the highest accuracy on the validation set (supervised training).

\subsection{Ablation study}
\label{ablation-study}
We have carried out an ablation study on the components of our model. 
Firstly, regarding the combination of the lateral connection, the first option (connection from the \textit{Regular} to the \textit{Fast} pathways and from the \textit{Fast} to the \textit{Slow} pathways) provided a better accuracy. Therefore, this is the one we will be using next.

\begin{table}[h!]
\centering
\begin{tabular}{|c|}
\hline
\textbf{SFR model} \\
\hline
\begin{tabular}{ll}
\textbf{Module} & \textbf{Acc} \\
\midrule
with 3D-Resnet18 & 72.9 \\
with 3D-Resnet50 & 71.6 \\
\end{tabular} \\
\hline
\end{tabular}
\caption{Results on the binary classification in terms of accuracy (\textit{Acc}) obtained by our SFR model with ResNet18 and ResNet50 modules using the cross entropy loss. We have carried out a dozen evaluations each time for different training seeds and folds, and we provide the average.
}
\label{tab:resnet50_resnet18}
\end{table}

We conducted an ablation experiment on the depth of the ResNet network to be used. We tested two possible depths, 18 layers and 50 layers, which would allow us to still have a rather light model. We trained our SFR model with ResNet18 and ResNet50 modules. For this experiment, we simply took the cross entropy loss, pending a decision on the $\gamma$ parameter of the focal loss. Results on the binary classification (transferable \textit{vs} non-transferable) are reported in Table \ref{tab:resnet50_resnet18}. Since ResNet-18 results are slightly better and it is even lighter, we selected this architecture.

We omitted the directed lateral connections between the three pathways of our SFR model. We call SFR Late Fusion the model without any lateral connections. We use ResNet18 as motivated above. In the same time, we investigated both the cross entropy loss and the focal loss, thus combining these two ablation experiments. For the focal loss, we took $\gamma=2$, as justified below.
We set $\alpha_1=1.25$ and $\alpha_2=0.833$ according to the frequency of the two classes. Results are reported in Table \ref{tab:connection-loss}. 
In all cases, the focal loss leads to increased performance. Furthermore, SFR performs better than SFR Late Fusion, which shows the importance of lateral connections.

\begin{table}[h]
\centering
\begin{tabular}{|c|}
\hline
\textbf{SFR and SFR Late Fusion models} \\
\hline
\begin{tabular}{ll}
\textbf{Model and loss function} & \textbf{Acc} \\
\midrule
SFR(CE) & 72.9 \\
SFR(FL) & 75.6 \\
SFR Late Fusion(CE) & 72.5 \\
SFR Late Fusion(FL) & 73.5 \\
\end{tabular} \\
\hline
\end{tabular}
\caption{Results  in terms of accuracy (\textit{Acc}) obtained by SFR and SFR Late Fusion with the cross entropy loss (CE) and focal loss (FL) for $\gamma=2$. We have carried out a dozen evaluations each time for different training seeds and folds, and we provide the average.
}
\label{tab:connection-loss}
\end{table}

\begin{table}[h!]
\centering
\begin{tabular}{|c|}
\hline
\textbf{SFR model} \\
\hline
\begin{tabular}{ll}
\textbf{Gamma value} & \textbf{Acc} \\
\midrule
SFR(FL) ($\gamma=1$) & 73.1 \\
SFR(FL) ($\gamma=2$) & 75.6 \\
SFR(FL) ($\gamma=3$) & 73.4 \\
\end{tabular} \\
\hline
\end{tabular}
\caption{Results on the binary classification in terms of accuracy (\textit{Acc}) obtained by our model SFR (involving ResNet18) with focal loss and $\gamma\in\{1,2,3\}$. We have carried out a dozen evaluations each time for different training seeds and folds, and we provide the average.
}
\label{tab:focal-loss}
\end{table}

Our last ablation experiment dealt with the setting of parameter $\gamma$ of the focal loss function. We tested the focal loss for three values of parameter $\gamma$, $\gamma=1,2, \; \text{and} \; 3$, knowing that the case $\gamma=0$ is somehow equivalent to train the model with a weighted binary cross-entropy loss.
We carried out the experiment on the binary classification with our SFR model. We report the results in Table \ref{tab:focal-loss}. We can conclude that the best choice is $\gamma=2$. By the way, the authors of \cite{human_embryo2} made the same choice.

\subsection{Comparative experiments}
We have carried out comprehensive comparative experiments on the early prediction of bovine embryo transferability. To evaluate the performance of all methods, we consider the following metrics: overall accuracy \textit{Acc}, precision $P_T$ (resp. $P_{NT}$) and recall $R_T$ (resp. $R_{NT}$) for the $T$ (resp. $NT$) class.
We performed the binary classification with SlowFast \cite{slowfast} (the ResNet18 version of the code) and a classical 3D-RestNet18, using for both the focal loss, since this loss is more adapted to our problem as demonstrated in Table \ref{tab:connection-loss}. We train the two methods on our training dataset. This yields a comparison between these two models and ours.
In addition, we built a baseline model comprising a 2D convolutional neural network (CNN) followed by a recurrent neural network (RNN). The 2D CNN is implemented with a ResNet18 that learns spatial features on images. A recurrent GRU neural network handles the time dimension of the embryo video, taking as input the successive output of the 2D CNN. The output of the last cell of the GRU is sent to a fully connected layer to obtain the classification prediction.

Comparative results for all the tested models are collected in Table \ref{table:different_architectures} with the use of the focal loss for all methods. As expected, the 2D network involving a recurrent neural network underperforms all the 3D networks. 
Our SFR method obtains the best accuracy rate, followed by ResNet18. In addition our model is much more stable than the others, on the different metrics, which is crucial.
SlowFast does not perform as expected, probably due to the particular nature of the videos processed, very different from those
videos considered in action recognition. In addition, our SFR method has the best precision score for the T class and the best recall score for the NT class, which is very important for the target application. Indeed, for cattle breeding, it is essential to predict transferable embryos correctly, in order to avoid unnecessary pregnancies by transferring non-transferable embryos.

\begin{table*}[h!]
\centering
\begin{tabular}{l|c||c|c|c|c|c}
\multicolumn{1}{c|}{Model} & $Acc$      & $P_T$      & $R_T$      & $P_{NT}$   & $R_{NT}$ & Average Prediction Time \\ \hline
2D-Resnet18+GRU&   67.6$\pm$4.1       &   54.6$\pm$5.6     &   48.4$\pm$19.9   &    74.2$\pm$5.9&   78.3$\pm$8.6 & 0.037s   \\
3D-Resnet18(FL)     &  74.6$\pm$4.0      &   64.1$\pm$6.5     &    67.4$\pm$12.4    &   82.0$\pm$4.6   & 78.6$\pm$7.8 & 0.205s\\
SlowFast(FL)        &  73.1$\pm$2.6       &   63.9$\pm$4.0    &  55.2$\pm$9.3      &   77.3$\pm$3.2    &    82.8$\pm$3.4 & 0.077s \\
Ours - SFR(FL) &  \textbf{75.6$\pm$1.5} & \textbf{66.8$\pm$3.3} & 62.5$\pm$3.4 & 80.1$\pm$1.1 &  \textbf{82.8$\pm$2.9} & 0.294s \\
\end{tabular}
\caption{Comparison of results obtained for the models 2D-Resnet18+GRU, 3D-Resnet18, SlowFast \cite{slowfast}, and our model SFR, all models with the focal loss ($\gamma=2$). We have carried out a dozen evaluations each time for different training seeds and folds, and we provide the mean and standard deviation for the accuracy, precision and recall for the $T$ and $NT$ classes.
}
\label{table:different_architectures}
\end{table*}

We have also compared our method with others regarding the computation time to make a prediction, still in Table \ref{table:different_architectures}. For each model, we computed the average time for one inference by repeating 1000 predictions on a NVIDIA RTX A500. As expected, the 2D CNN with GRU is the fastest at inference, followed by SlowFast which is faster than a simple 3D-ResNet as shown in the original paper\cite{slowfast}. Next come 3D-ResNet18 and our SFR model, whose average time prediction is increased by approximately $89 ms$ compared to the 3D-ResNet18, but this is not a problem for our application.

\begin{figure}[tb!]
\centering
\resizebox{\columnwidth}{!}{
        \includegraphics{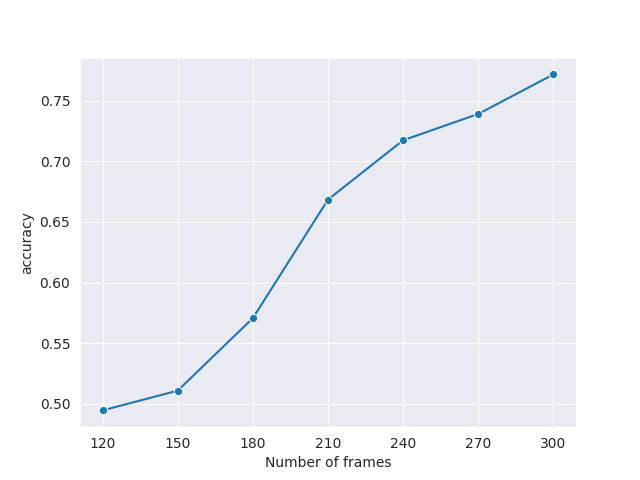}   
}
\caption{Accuracy of early prediction for our SFR model with focal loss for the following range of video length: from 120 frames (about two days 1/4 of embryo development) to 300 frames (about four days).}
\label{fig:early_prediction}
\end{figure}

\subsection{Earlier prediction}
We wanted to check whether it is possible to make a prediction even before four days have elapsed. Therefore, we evaluate the performance of our SFR model with focal loss, when performing less-than-4-day Transferable \textit{vs} Non transferable prediction. To do this, we test the model with increasingly shorter videos, removing the last thirty frames each time, which corresponds to removing information occurring during seven hours and a half. We stop testing at 120 frames, i.e., approximately two days 1/4 of embryo development.
Results are plotted in Fig.\ref{fig:early_prediction}. We observe that the curve regularly climbs.
Depending on the accuracy level acceptable for a given application, a usable prediction could be provided at an even earlier stage than the four days of the embryo development.

\section{Conclusion}
\label{conclusion}

We have designed a 3D model to predict the embryo transferability within four days, taking 2D time-lapse microscopy videos as input. We state the problem as a supervised two-class classification one that remains however difficult. The three-pathway architecture makes our 3D model multi-scale in time regarding the input videos and able to manage appearance and motion in different ways. We successfully dealt with
poorly discriminating embryo appearance and motion, in addition affected by transparency, and small inter-class distance. Experiments demonstrate the usefulness of directed lateral connections between pathways, and of the focal loss. We favorably compared our SFR model with other methods. SFR provides the best accuracy rates with much better stability than 3D-ResNet18 on all metrics. Thus, we are able to efficiently and accurately achieve the early prediction of bovine embryo transferability. Future work will be concerned with an explicit combination of classification earliness and accuracy in the training loss.


\section{Compliance with Ethical Standards}
This research study was conducted using data available in our laboratory. No live animals or euthanised animals were used to create the original data. The semen was acquired from a commercial company and the cumulus oocyte complex were harvested from ovaries recovered \textit{post-mortem} in a commercial slaughterhouse. Both these companies and our laboratory are based in France and state-approved. The necessary authorisations for the use of \textit{post-mortem} biological material have been obtained from the responsible Ministry.

\section{Acknowledgements}
The authors would like to acknowledge the collaboration of Dr. Véronique Duranthon and Brigitte Marquant-LeGuienne for the experimental protocol design for the embryo production.
The authors declare no conflicts of interest.
The production of the original embryo data was funded by CRB-Anim.

Yasmine Hachani's PhD grant is funded by Inria. Operation of the research project is also partly funded by the DIGIT-BIO program of INRAe.

\bibliographystyle{IEEEbib}

\end{document}